\documentclass[%
 aip,
 amsmath,amssymb,
 reprint,%
]{revtex4-1}

\usepackage{graphicx}
\usepackage{dcolumn}
\usepackage{bm}
\usepackage[dvipsnames]{xcolor}

\usepackage[utf8]{inputenc}
\usepackage[T1]{fontenc}
\usepackage{mathptmx}
\usepackage[english]{babel}
\usepackage{etoolbox}
\usepackage{booktabs} 
\usepackage{multirow} 

\makeatletter
\def\@email#1#2{%
 \endgroup
 \patchcmd{\titleblock@produce}
  {\frontmatter@RRAPformat}
  {\frontmatter@RRAPformat{\produce@RRAP{*#1\href{mailto:#2}{#2}}}\frontmatter@RRAPformat}
  {}{}
}%
\makeatother
\begin{document}
\selectlanguage{english}

\preprint{AIP/123-QED}

\title{Second-Order Optical Nonlinearity of AlScN Films Grown By Molecular Beam Epitaxy}
\author{Joongwon Lee}
\affiliation{School of Electrical and Computer Engineering, Cornell University, Ithaca, New York 14853, USA}
\email{jl3755@cornell.edu}

\author{Thai-Son Nguyen}
\affiliation{Department of Materials Science and Engineering, Cornell University, Ithaca, New York 14853, USA}

\author{Len van Deurzen}
\affiliation{Department of Applied and Engineering Physics, Cornell University, Ithaca, New York 14853, USA}

\author{Debaditya Bhattacharya}
\affiliation{School of Electrical and Computer Engineering, Cornell University, Ithaca, New York 14853, USA}

\author{Chandrashekhar Savant}
\affiliation{Department of Materials Science and Engineering, Cornell University, Ithaca, New York 14853, USA}

\author{Siddhartha Ghosh}
\affiliation{Northrop Grumman Corporation, Redondo Beach, CA 90277, USA}

\author{Patrick Shea}
\affiliation{Northrop Grumman Corporation, Linthicum, MD 21090, USA}

\author{Carl Bernard}
\affiliation{School of Electrical and Computer Engineering, Cornell University, Ithaca, New York 14853, USA}

\author{Huili Grace Xing}
\affiliation{School of Electrical and Computer Engineering, Cornell University, Ithaca, New York 14853, USA}
\affiliation{Department of Materials Science and Engineering, Cornell University, Ithaca, New York 14853, USA}

\author{Debdeep Jena}
\affiliation{School of Electrical and Computer Engineering, Cornell University, Ithaca, New York 14853, USA}
\affiliation{Department of Materials Science and Engineering, Cornell University, Ithaca, New York 14853, USA}

\author{Farhan Rana}
\affiliation{School of Electrical and Computer Engineering, Cornell University, Ithaca, New York 14853, USA}

\date{\today}

\begin{abstract}
Alloys of AlN have rapidly emerged as a material platform for nonlinear optics. In this paper, we measure the second-order optical nonlinearity of AlScN films grown directly on nitrided c-plane sapphire by molecular beam epitaxy. This direct growth approach, which bypasses a thick AlN buffer layer, allows us to isolate the true nonlinear response of the AlScN film. Our results show a large enhancement of $d_{31}$, but a suppression of $d_{33}$ in AlScN films compared to AlN. We observe that $d_{31}$ can be as high as 4.92 $pm/V$, which is 60 times larger than that of AlN. The development of AlScN-based photonic devices can enable energy-efficient nonlinear optical operations that can be epitaxially integrated with electronic and photonic devices based on Si, GaN and AlN.
\end{abstract}

\maketitle

Silicon-on-insulator~\cite{Jalali2006,Leuthold2010,Lim2014} and lithium niobate (LiNbO$_3$)~\cite{Boes2023} platforms have long driven integrated optics, yet they face persistent limitations in centrosymmetry constraints and CMOS compatibility, respectively. Aluminum scandium nitride (AlScN) has attracted significant interest as a potential alternative for on-chip nonlinear photonic devices due to its structural compatibility with GaN and Si semiconductor platforms~\cite{Jena2019,Kim2023,Casamento2021,Zhang2023} and its tunable material properties~\cite{Akiyama2009APL, Akiyama2009Advm, casamento2023alscn, suceava2023,casamento2022epitaxial, hayden2021ferroelectricity, gremmel2025effect,  savant2025epitaxial,Fichtner2019, casamento2022ferrohemts, Savant2024APL, Savant2024PSS}. Because it can be integrated directly with existing semiconductor workflows, understanding its second-order optical nonlinearity ($\chi^{(2)}$) is essential for evaluating its viability in compact, low-power components performing functionalities such as second harmonic generation (SHG), optical parametric amplification (OPA), and electro-optic modulation (EOM). In these photonic applications, the longitudinal nonlinear coefficient (d$_{33}$) is of primary importance, as it typically dictates the maximum conversion efficiency for standard waveguide geometries and polarization modes.

However, recent literature presents conflicting accounts regarding the strength of this critical coefficient. While initial measurements on sputtered AlScN on Ti/Pt reported an exceptionally large d$_{33}$ value up to 62.3 $pm/V$~\cite{Yoshioka2021}, a recent study on molecular beam epitaxy (MBE)-grown AlScN on AlN/sapphire suggests that d$_{33}$ decreases toward zero at higher scandium compositions, accompanied by an increase in d$_{31}$~\cite{Theriault2026}. Resolving this discrepancy is crucial for the realistic design of AlScN-based photonic components. In this article, we address this conflict by characterizing AlScN thin films grown directly on nitrided sapphire substrates via MBE, reporting the full second-order nonlinear optical susceptibility tensors. We further address whether the 150 nm thick buffer AlN layer used in the recent study introduces a difficulty in isolation of the active AlScN layer in the analysis.

\begin{figure*}
\includegraphics[width=\textwidth]{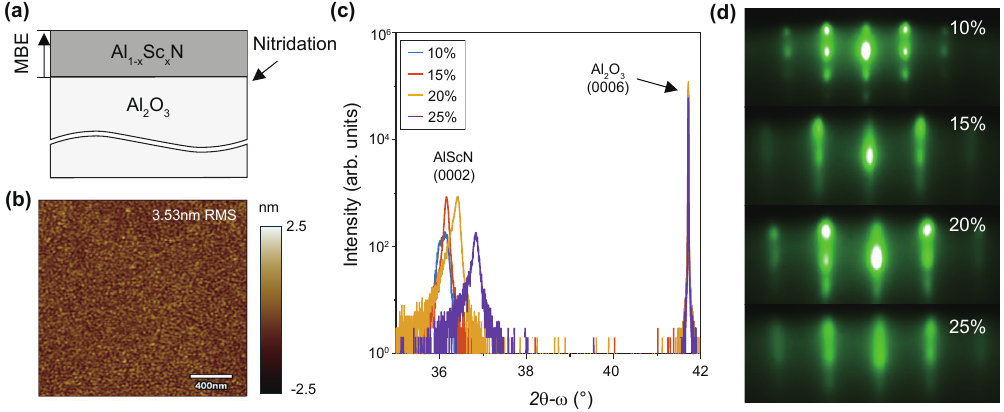}
\caption{\label{Fig_AlScN_Struc} (a) Heterostructure schematic of MBE-grown-AlScN films on nitrided sapphire. (b) Atomic force microscopy (AFM) scan of an as-grown 20\% Sc containing AlScN film sample. The root mean square (RMS) surface roughness is 0.353 nm. AFM scans for other samples are provided in Fig. S1. (c) X-ray diffraction (XRD) symmetric $2\theta-\omega$ scans for the 10\%, 15\%, 20\%, 25\% Sc containing AlScN film samples. (d) Reflection high energy electron diffraction (RHEED) patterns for the 10\%, 15\%, 20\%, 25\% Sc composition samples.}
\end{figure*}

The AlScN samples in this study were grown on c-plane sapphire substrates with a 2-inch diameter using plasma-assisted MBE. First, the sapphire substrate was subjected to a surface nitridation process, which has been reported to enhance the lateral heteroepitaxy of group III-nitride on sapphire~\cite{namkoong_role_2002}. Crucially, this process bypasses a thick AlN buffer layer conventionally used in MBE growth. By keeping the interfacial AlN layer to the ultrathin limit, we eliminate the parasitic optical interference and potential SHG from the buffer. The nitridation was performed at a substrate thermocouple temperature of 200 $^\circ$C for three hours using nitrogen plasma with a flow rate of 1.95 sccm and an RF plasma power of 400 W. Subsequently, four AlScN films with target thickness of 150 nm were grown with compositions of 10, 15, 20, and 25\% Sc at a substrate thermocouple temperature of 530 $^\circ$C using 1.95 sccm nitrogen flow with 200 W RF plasma power and a III/V ratio of 0.7. The resulting MBE heterostructure is shown in Fig.~\ref{Fig_AlScN_Struc}(a). The detailed growth conditions of MBE AlScN have been reported elsewhere\cite{nguyen_lattice-matched_2024}. We monitored the AlScN growth in situ using a kSA Instruments reflection high energy electron diffraction (RHEED) apparatus with a Staib electron gun operating at 14.5 kV and 1.45 A.

The MBE-grown AlScN samples were characterized using the following techniques and instruments. We used a Panalytical Empyrean system with Cu K$_{\alpha1}$ radiation to perform X-ray diffraction (XRD) measurement on the thin films. The surface morphology was characterized using an Asylum MFP-3D ES atomic force microscope (AFM). We determined the Sc composition by X-ray photoelectron spectroscopy (XPS) using a Scienta-Omicron-ESCA-2SR XPS instrument equipped with a 1486.6 eV Al K$_{\alpha}$.

AFM scans confirm the presence of island-like features. The lowest RMS roughness of 0.353 nm was obtained for the 20\% Sc containing AlScN film (Fig. \ref{Fig_AlScN_Struc}(b)), and the highest RMS roughness of 1.370 nm was obtained for the 10\% Sc AlScN (see Fig. S1 for AFM scans for all films). These results further corroborate the shift in RHEED patterns with varying Sc composition. The crystalline quality and surface morphology of MBE-grown AlScN films can be further improved in future studies by optimizing growth temperature, III/V ratio, and strain management by adding a buffer nucleation layer\cite{hardy_control_2020} and optimizing sapphire nitridation conditions\cite{namkoong_role_2002}.

The symmetric 2$\theta/\omega$ XRD scans in Fig.~\ref{Fig_AlScN_Struc}(c) show only the AlScN 0002 and Al$_2$O$_3$ 0006 Bragg reflections, suggesting that epitaxial wurtzite AlScN was achieved without the formation of additional intermetallic phases\cite{hardy_epitaxial_2017}. By normalizing the substrate Al$_2$O$_3$ peaks, we find that the 15\% and 20\% Sc containing AlScN films exhibit stronger c-axis orientation. The increase in XRD intensity of the AlScN 0002 peak from the 10\% Sc to 20\% Sc containing films can be attributed to an increased 2D growth mode with increasing Sc composition, as reported by Hardy et al.\cite{hardy_control_2020}, which enables high crystalline quality and strong c-axis orientation. The weaker XRD intensity of the 25\% Sc AlScN 0002 peak is likely due to an increase in lattice distortion as high Sc AlScN films undergo a transition into a wurtzite crystal with zinc blende or rocksalt phase inclusions\cite{hardy_epitaxial_2017,hardy_control_2020, casamento_structural_2020}. Similar XRD peak position, intensity, and sharpness trends have been reported for both sputtered and MBE AlScN~\cite{Yoshioka2021,WangP2020}.

Fig.~\ref{Fig_AlScN_Struc}(d) shows the spot-modulated streak pattern observed of the AlScN films after growth, suggesting a Stranski–Krastanov growth mechanism with layer-plus-island formation due to the N-rich growth condition of AlScN\cite{nguyen_lattice-matched_2024}. As the Sc composition increases from 10\% to 20\%, we observe a transition from a spottier (3D growth) to a more spot-modulated streak pattern (mixed 2D-3D growth). The RHEED pattern becomes more diffuse for the 25\% Sc sample, implying a lower crystalline quality compared to the 15\% and 20\% Sc samples. These trends observed in RHEED agree with the changes in symmetric XRD peak intensity and sharpness observed in Fig.~\ref{Fig_AlScN_Struc}(c).

To minimize scattering losses during transmission-mode optical measurements, we prepared the back surface (non-epitaxial side) of the sapphire substrate via a multi-stage mechanical polishing process. We employed a sequence of diamond lapping films with progressively finer grit sizes, ranging from 30 $\mu$m down to 0.1 $\mu$m. The rotational speed was graduated to match the grit size, starting at 100 rpm for the coarsest films and progressively decreasing to 75 rpm for the final polishing stages. Each step lasted approximately 30 to 90 minutes, ensuring an optically smooth interface for the transmission geometry.

\begin{figure*}
\includegraphics[width=\textwidth]{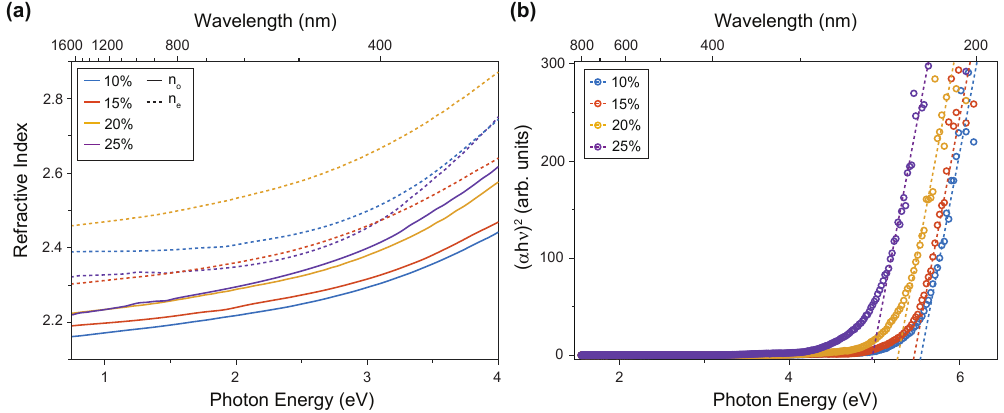}
\caption{\label{Fig_LO_Prop}(a) Dispersion (ordinary and extraordinary refractive index) measured by ellipsometry, and (b) Tauc plots measured for the 10\%, 15\%, 20\%, 25\% Sc containing AlScN film samples.}
\end{figure*}

We first characterized the linear optical properties of the AlScN films, as accurate linear constants are essential for determining nonlinear optical susceptibilities. Fig.~\ref{Fig_LO_Prop}(a) illustrates the dispersion of the refractive index measured by spectroscopic ellipsometry. To correctly model the behavior of p-polarized light in the uniaxial (0001)-oriented AlScN crystal, a birefringent model was employed to fit the ellipsometric data, modeling both the ordinary (n$_o$) and extraordinary (n$_e$) index. The thicknesses of the films were determined to be 136 nm, 145 nm, 133 nm, and 170 nm for the 10\%, 15\%, 20\%, 25\% Sc containing AlScN film samples, respectively.

For the 10$\%$ film, n$_o$ increases from 2.16 at 1600 nm to 2.4 at 300 nm. Consistent with previous reports\cite{vanDeurzen2023}, n$_o$ increases with Sc concentration; the 25$\%$ film exhibits an index approximately 0.06 to 0.2 larger than the 10$\%$ film, depending on the wavelength. Furthermore, we measure n$_e$ to be approximately 0.3 larger than n$_o$. Notably, spectroscopic ellipsometry exhibits limited sensitivity to n$_e$ - a consequence of constrained internal propagation angles - which precludes a reliable assessment of n$_e$ scaling with Sc concentration. Despite these limitations, our optical characterization provides a robust foundation for interpreting the subsequent nonlinear optical measurements. 

We further confirm that interband absorption does not contribute significantly to the optical response at our operating wavelength by measuring the optical bandgap of each sample. Absorbance spectra were acquired using a Varian Cary 500 Scan UV-VIS NIR Spectrophotometer. Fig.~\ref{Fig_LO_Prop}(b) shows the Tauc plot measured for each sample. For all films, $(\alpha h \nu)^2$ remains negligible for lower photon energies and increases linearly after the bandgap, which is consistent with a direct bandgap model of AlScN. The bandgap was extracted by identifying the intercept of the linear region of $(\alpha h \nu)^2$ vs. $h \nu$ plot. The bandgap decreases monotonically from approximately 5.5 eV at 10$\%$ Sc composition to 5 eV at 25$\%$ Sc composition, consistent with previous reports~\cite{Baeumler2019}. The extracted bandgap values for all AlScN films are significantly higher than the SHG photon energies used in our measurements (2.40 eV), ensuring that the measurement remains in the transparent regime of the material.

\begin{figure*}
\includegraphics[width=\textwidth]{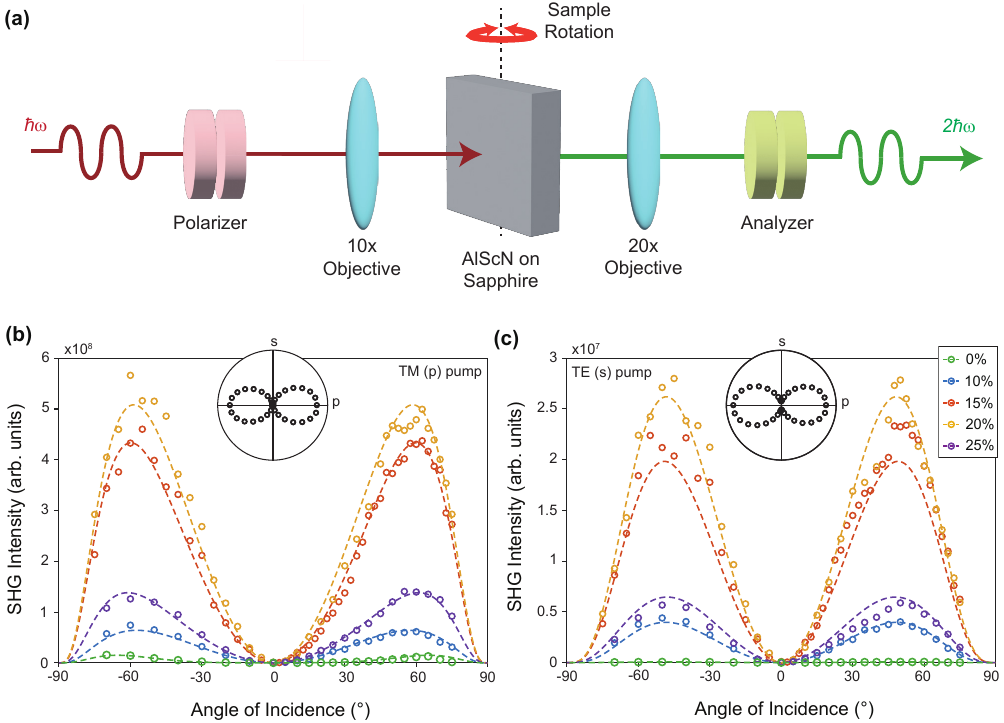}
\caption{\label{Fig_NLO_Meas}(a) Experimental scheme depicting the optical setup geometry for the SHG measurements. (b), (c) Measured p-polarized SHG intensity as a function of angle of pump incidence for p-polarized (b) and s-polarized pump (c). Inset: Polarization dependence of the SHG intensity.}
\end{figure*}

We next evaluate the second-order nonlinear optical properties of the AlScN films. AlScN crystallizes in a wurtzite crystal structure (space group $P6_3mc$), which dictates the form of the second-order susceptibility tensor:
\begin{equation} \label{Eq_Wurtzite_SHG}
    \chi^{(2)}=2
    \begin{pmatrix}0 & 0 & 0 & 0 & d_{31} & 0 \cr
    0 & 0 & 0 & d_{31} & 0 & 0 \cr
    d_{31} & d_{31} & d_{33} & 0 & 0 & 0 \cr
    \end{pmatrix}.
\end{equation}
To decouple and determine the individual magnitudes of $d_{31}$ and $d_{33}$, we performed SHG measurement as a function of angle of incidence~\cite{Bloembergen1962, Kiehne1998, Miragliotta1993}. The experimental configuration is illustrated in Fig.~\ref{Fig_NLO_Meas}(a). The samples were illuminated with a pulsed laser (1032 nm wavelength, 150 fs pulse width, 100 MHz repetition rate) focused at a 50 $\mu$m diameter spot size using a 10$\times$ objective lens, resulting in a fluence of approximately 20 $ \mu$J / cm$^2$. A polarizer and an analyzer were used to precisely control the polarization states of the pump beam and the detected SHG signals. The resulting SHG signals were collected in a transmission geometry. The angle of incidence was precisely controlled using a motorized rotation stage, and the SHG intensity was recorded using a Horiba iHR-320 monochromator with a Synapse charge-coupled device (CCD). 

Fig.~\ref{Fig_NLO_Meas} (b) and (c) display the measured SHG intensity as a function of the incident angle for p-polarized and s-polarized pump beams, respectively. For all measurements, only the p-polarized component of the SHG signal was recorded, as polarization-resolved measurements confirmed the signal to be purely p-polarized SHG (see insets of Fig.~\ref{Fig_NLO_Meas} (b), (c)). This result is consistent with the symmetry requirements of the $P6_3mc$ space group. Specifically, for a p-polarized pump, ($\vec{E}(\omega)=E_x\hat{x}+E_z\hat{z}$), the nonlinear polarization is:
\begin{equation}
    \vec{P}(2\omega)=2d_{31}E_xE_z\hat{x}+(d_{31}E_x^2+d_{33}E_z^2)\hat{z},
\end{equation}
which resides entirely in the x-z (incidence) plane and is thus p-polarized. Similarly, for an s-polarized pump ($\vec{E}=E_y\hat{y}$), the nonlinear polarization is:
\begin{equation}
    \vec{P}(2\omega)=d_{31}E_y^2\hat{z},
\end{equation}
which is also p-polarized.

The SHG intensity generated by a p-polarized pump is more than one order of magnitude larger than that produced by an s-polarized pump. While the SHG intensity vanishes for normal incidence in both configurations, the p-polarized response peaks at 60$^\circ$, whereas the s-polarized response peaks at 50$^\circ$. We observed an initial increase in the SHG intensity with the concentration of Sc. However, the 25$\%$ sample exhibits a significantly decreased SHG intensity compared to the 15$\%$ and 20$\%$ samples. To quantify these results and extract the individual $d_{31}$ and $d_{33}$ coefficients, we employed a transfer matrix method (TMM) approach specifically adapted for our SHG geometry. The dashed lines in Fig.~\ref{Fig_NLO_Meas} (b) and (c) represent the TMM simulations, which shows excellent agreement with the experimental data, accurately reproducing the angular dependence of SHG signals for both s- and p-polarized pump.

\begin{figure*}
\includegraphics[width=\textwidth]{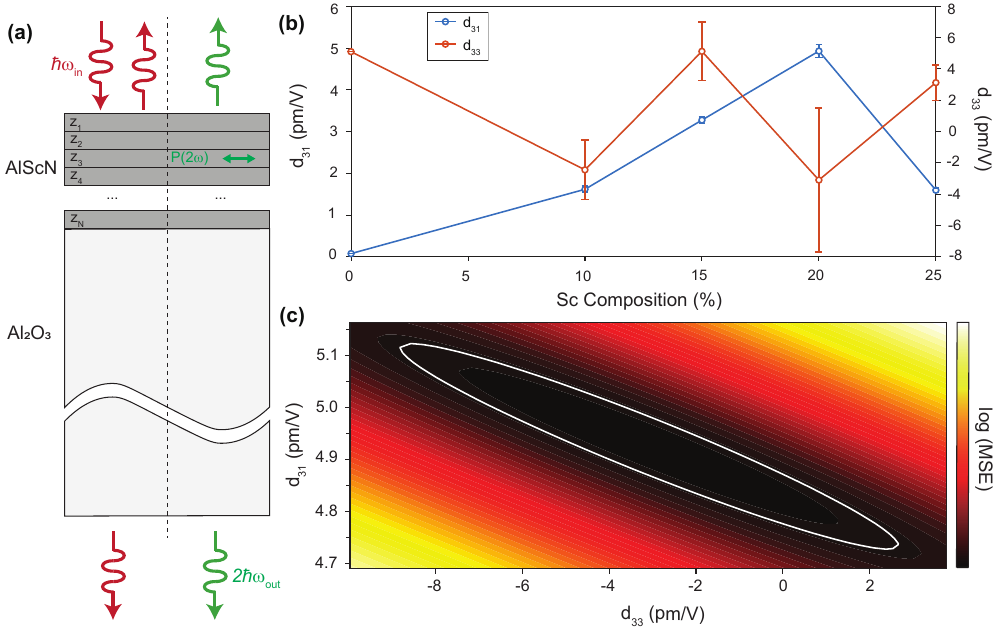}
\caption{\label{Fig_TMM_Fit}(a) Schematic describing the transfer matrix method (TMM) simulations. (b) $d_{31}$ and $d_{33}$ plotted for various Sc compositions. (c) Mean squared error (MSE) in fitting the 20\% Sc film data for various $d_{31}$ and $d_{33}$ values.}
\end{figure*}

The TMM simulation model is implemented as follows~\cite{Bethune1989}. For a given angle of incidence ($\theta_i$), the pump electric field vector $\vec{E}(\omega,z)$ is computed throughout the AlScN layer using a linear TMM (Fig.~\ref{Fig_TMM_Fit}(a), left). The layer is then discretized into thin slices of thickness $\Delta z$. At each position $z_i$, the nonlinear polarization $\vec{P}(2\omega,z_i)$ acts as a local source for the harmonic field (Fig.~\ref{Fig_TMM_Fit}(a), right). In this framework, the polarization introduces a discontinuity in the transverse electric ($E_x$) and magnetic ($H_y$) field components across each computational slice, defined as:
\begin{equation} \label{Eq_SE}
    S_E=-i\frac{k_x}{\varepsilon_0\varepsilon_{zz}}P_z\Delta z,
\end{equation}
\begin{equation}
    S_H=i\omega P_x\Delta z,
\end{equation}
where $k_x$ is the transverse wavevector, $\varepsilon_0$ is the vacuum permittivity, $\varepsilon_{zz}$ is the extraordinary permittivity at the SHG wavelength, $\omega$ is the angular frequency of the SHG light. The total SHG signal is determined by summing the contributions from all slices propagated to the sapphire substrate. For these calculations, the substrate is treated as a semi-infinite medium as its thickness exceeds the coherence length of the laser pulses. Finally, the transmission through the sapphire/air interface is accounted for to determine the total emitted SHG intensity. This approach inherently fully incorporates phase-mismatch and multiple reflection effects.

To calibrate the experimental setup and validate the TMM analysis, we initially measured an AlN (0$\%$ Sc) control sample. Our results were benchmarked against the literature values of $d_{31}=$0.08 $pm/V$, $d_{33}=$5.1 $pm/V$~\cite{Fujii1977,Majkic2017,Larciprete2006,Yoshioka2021}, with the $d_{31}/d_{33}$ ratio slightly adjusted to optimize the fit to our baseline data. Following calibration, the second order nonlinearities of the AlScN films were determined across the range of Sc compositions. The magnitudes of $d_{31}$ and $d_{33}$ for each sample were extracted by simultaneously minimizing the mean squared error (MSE) between the TMM simulations and the experimental data for both the p-polarized (Fig.~\ref{Fig_NLO_Meas}(b)) and s-polarized (Fig.~\ref{Fig_NLO_Meas}(c)) configurations. The error represents the 95$\%$ confidence interval of the fit.

Fig.~\ref{Fig_TMM_Fit}(b) summarizes the extracted $d_{31}$ and $d_{33}$ values for AlScN as a function of Sc concentration. Our results show that that while $d_{31}$ increases significantly from 0.08 $pm/V$ at 0$\%$ Sc to approximately 5$pm/V$ at 20$\%$ Sc, followed by a decrease at 25$\%$ Sc. In contrast, $d_{33}$ appears unenhanced and is characterized by a high degree of uncertainty. This discrepancy stems from the high refractive index of AlScN; due to refraction at the air-film interface, the pump beam propagates at a relatively small internal angle. Consequently, the longitudinal component of the nonlinear polarization ($P_z$) is suppressed by a factor of $\varepsilon_{zz}$ (see Eq.~\ref{Eq_SE}), making the SHG intensity far less sensitive to variations in $d_{33}$ than $d_{31}$.

This sensitivity imbalance is further illustrated in the error analysis for the 20$\%$ sample (Fig.~\ref{Fig_TMM_Fit}(c), see Fig. S2 for MSE plots for all films). The logarithmic MSE plot reveals a 95$\%$ confidence interval (white contour) with two distinct features: (i) an elongated elliptical shape, indicating a statistical trade-off between $d_{31}$ and $d_{33}$ during the fitting process, and (ii) a dramatic difference in precision. While $d_{31}$ is tightly constrained between 4.73 $pm/V$ to 5.12 $pm/V$ in $d_{31}$, $d_{33}$ spans a much broader range from -8.85 $pm/V$ to 2.66 $pm/V$, confirming that $d_{31}$ is the dominant coefficient extracted from this transmission geometry.

\begin{table}[htbp]
\centering
\begin{tabular}{lcccccc}
\toprule
\multirow{1}{*}{\textbf{Growth Method}} & \textbf{Sc (\%)} & \multicolumn{1}{c}{\textbf{$d_{31}$ (pm/V)}} & \multicolumn{1}{c}{\textbf{$d_{33}$ (pm/V)}} \\
\midrule
\multirow{5}{*}{MBE (This Work)} 
 & 0  & 0.08 & 5.1 \\
 & 10 & 1.61$\pm$0.07 & -2.45$\pm$1.89 \\
 & 15 & 3.27$\pm$0.08 & 5.12$\pm$1.87 \\
 & 20 & 4.93$\pm$0.16 & -3.10$\pm$4.59 \\
 & 25 & 1.60$\pm$0.05 & 3.21$\pm$1.15 \\
\midrule
\multirow{4}{*}{Sputter (Ref.[~\cite{Yoshioka2021}])} 
 & 0  & 0.07$\pm$0.006 & 5.1$\pm$0.4 \\
 & 10 & 0.84$\pm$0.05 & 15.8$\pm$1.2 \\
 & 20 & 2.4$\pm$0.1 & 42.5$\pm$3.3 \\
 & 28 & 3.4$\pm$0.2 & 46.9$\pm$3.9 \\
 & 36 & 4.5$\pm$0.3 & 62.3$\pm$5.6 \\
\midrule
\multirow{4}{*}{MBE (Ref.[~\cite{Theriault2026}])} 
 & 0  & 0.15$\pm$0.02 & 3.2$\pm$0.3 \\
 & 10 & 0.84$\pm$0.05 & 1.3$\pm$0.1 \\
 & 20 & 1.81$\pm$0.03 & 0.0$\pm$3.5 \\
 & 30 & 2.59$\pm$0.04 & 0.3$\pm$1.3 \\
\bottomrule
\end{tabular}
\caption{\label{Tab_AlXN_Properties}Summary of optical nonlinearity and bandgap of MBE-grown and sputtered AlScN films~\cite{Yoshioka2021, Theriault2026, Baeumler2019}.}
\end{table}

Nevertheless, the enhancement of $d_{31}$ is remarkably large, considering the wide bandgap of AlScN. Table.~\ref{Tab_AlXN_Properties} summarizes the optical nonlinearity measured from AlScN in this work and Ref.[~\cite{Yoshioka2021, Baeumler2019,Theriault2026}]. The enhancement of piezoelectric coefficient in AlScN~\cite{Akiyama2009APL, Akiyama2009Advm} with the inclusion of ScN, which tends to stabilize in a rocksalt structure, causes lattice distortion in which the Sc-N bond is longer than the Al-N bond. This has been observed in sputtered AlScN and AlBN~\cite{Yoshioka2021,suceava2023enhancement}. While we expected to observe an increasing $d_{31}$ in AlScN films with up to 40\% Sc composition, the limit at which AlScN retains its wurtzite crystal structure, we observed a decrease starting at 25\% Sc. This decrease is likely due to the lattice distortion caused by high Sc composition and the lattice mismatch between the film and the sapphire substrate ~\cite{casamento_structural_2020}, which results in degraded film quality and inclusion of rock salt or zinc blende phases. This is consistent with the observation made in the weaker XRD intensity and a more diffuse RHEED pattern (See Fig.~\ref{Fig_AlScN_Struc}(c), (d)).

On the other hand, we do not observe a noticeable enhancement in $d_{33}$. Despite the relative insensitivity of the measurement scheme to $d_{33}$ in a high index film, a significant enhancement in $d_{33}$ would still have been captured. Rather, it appears suppressed within the large error margin, consistent with the most recent reports~\cite{Theriault2026, Yang2024}. By using an MBE AlScN layer directly grown on nitrided sapphire, we eliminate parasitic scattering and complex internal reflections. This direct-growth approach ensures that our observation of the pronounced $d_{31}$ enhancement and the suppressed $d_{33}$ is strictly isolated to the AlScN film.

We note that our results present a notable contrast to Ref. [\cite{Yoshioka2021}], which reported a significant enhancement in $d_{33}$ reaching 64 pm/V at 36\% Sc composition. This discrepancy between films deposited by distinct techniques may stem from differences in structure and defect density. For sputtered films deposited on metallic layers, strain is typically relaxed through the formation of domains with varying in-plane orientations. In contrast, during our MBE growth, strain relaxation primarily occurs through crack formation, which becomes increasingly evident at higher Sc compositions (see Fig. S3). This process introduces a higher density of cracks and associated point defects in our MBE-grown samples, potentially leading to optical absorption and scattering that reduces the nonlinear optical response. Another possible difference may lie in the analysis method - while the TMM inherently accounts for the longitudinal dielectric screening that suppresses the radiation from the out-of-plane $P_z$ component in a high-index film, simplified ray-optic methods may fail to fully capture this, leading to an overestimation of $d_{33}$.

In conclusion, we measure and report the second-order nonlinear optical susceptibility tensor values of AlScN films grown by MBE. Our results show that AlScN is a promising material system for realizing nonlinear optical devices due to high optical nonlinearity, wide optical bandgap, and compatibility with existing nitride-based material platforms, such as GaN and AlN~\cite{vanDeurzen2023, YangG2024}. Additionally, we highlight that our measurement technique offers a rapid metrology method for characterizing emerging nonlinear optical materials, including other AlN alloys, especially those grown by MBE on transparent substrates. This is important for nonlinear photonic applications, as such films need to be grown on materials with low optical loss.

\vspace{2mm}
\begin{acknowledgments}
\vspace{-2mm}
This work was supported in part by SUPREME, one of seven centers in JUMP 2.0, a Semiconductor Research Corporation (SRC) program sponsored by DARPA (characterization).  Additional support from Northrop Grumman is acknowledged.  The authors acknowledge the use of the Cornell NanoScale Facility (CNF), a member of the National Nanotechnology Coordinated Infrastructure (NNCI), which is supported by the National Science Foundation (NSF Grant NNCI-2025233).
\end{acknowledgments}
\vspace{-2mm}

\section*{Author Declarations}
\subsection*{Conflict of Interest}
The authors have no conflicts to disclose.

\section*{Data Availability}
The data that support the findings of this study are available from the corresponding author upon reasonable request.

\bibliography{arxiv_Reference}

\end{document}